\documentstyle[12pt]{article}
\begin{document}
\title{Determination of the angle $\gamma$ from nonleptonic $B_c \to
D_s D^0$ decays}
\author{A. K. Giri$^1$, R. Mohanta$^2$ and M. P. Khanna$^1$ \\
{\it $^1$ Physics Department, Panjab University, Chandigarh-160 014,
India}\\ {\it $^2$ School of Physics, University of Hyderabad, 
Hyderabad-500 046, India }}
\maketitle
\begin{abstract}
We note that the two body nonleptonic pure tree decays $B_c^\pm
\to D_s^\pm D^0(\bar D^0)$ and the corresponding vector-vector modes 
$B_c^\pm \to D_s^{* \pm } D^{*0}(\bar D^{* 0}) $ are  well
suited to extract the weak phase $\gamma$ of the unitarity triangle.
The CP violating phase $\gamma$ can
be determined cleanly as these decay modes are free from the 
penguin pollutions.
\end{abstract}

\section{Introduction}

Despite many attempts, CP violation still remains one of the
most outstanding problems in particle physics \cite{kog89,bigi00}.
The standard model (SM) with three generations provides a simple
description of this phenomenon through the complex
Cabibbo-Kabayashi-Maskawa matrix \cite{ref1}. Decays of $B$ mesons
provide rich ground for investigating $CP$ violation
\cite{buras98,quin98}. They allow  stringent tests both for the 
SM and for studies
of new sources of this effect. Within the SM, the $CP$ violation is
often characterized by the so-called unitarity triangle \cite{chau88}.
Detection of CP violation and the accurate determination of the
unitarity triangle are the major goals of experimental $B$
physics \cite{stone94}. Decisive information about the origin
of CP violation in the flavor sector can  be obtained if the three angles
$\alpha (\equiv \phi_2)$, $\beta(\equiv \phi_1)$
and $\gamma(\equiv \phi_3)$ can be independently measured
\cite{bigi91}. The sum of these three angles must be equal to
$180^\circ $
if the SM with three generations is the model for the $CP$ violation.
These angles of the unitarity triangle
can be loosely bounded from various low energy phenomenology. The angle
$\beta(\equiv \phi_1)$ can be measured from the gold plated mode
$B_d \to J/\psi K_s $ without any hadronic uncertainty. The angle
$\alpha (\equiv \phi_2)$ can be measured from $B \to \pi \pi $ mode but
there is some penguin contamination. Still one can hope to perform the
isospin analysis and remove the penguin contribution thereby 
extracting the angle $\alpha (\phi_2)$ with reasonable accuracy
\cite{ref2}. The most difficult to measure is the angle $\gamma (\equiv
\phi_3)$. There have been a lot of suggestions and discussions about
how to measure this quantity at $B$ factories \cite{ref3,ref4}. In Ref.
\cite{ref3} the authors proposed to extract $\gamma$ using the independent
measurements of $B \to D^0 K$, $B \to \bar D^0 K$ and $B \to D_{CP} K$.
However, for the charged B meson decay mode $A(B^- \to \bar D^0 K^-)$ is
difficult to measure experimentally. The reason is that the final
$\bar D^0$ meson should be identified using $\bar D^0 \to K^+ \pi^-$
but it is difficult to distinguish it from doubly Cabibbo suppressed
$D^0 \to K^+ \pi^-$.
There are various methods to overcome these difficulties. In
Ref. \cite{ref5} Atwood et al
used different final states into which the
neutral $D$ meson decays, to extract information on $\gamma$. In Ref.
\cite{ref6} Gronau proposed that the angle $\gamma$ can be determined
by using the color allowed decay modes $B^- \to D^0 K^-$, $B^- \to
D_{CP} K^- $ and their charge conjugation modes.
In Ref. \cite{ref7} a new method, using the isospin relations, 
is suggested to extract $\gamma$
by exploiting the decay modes $B \to D K^{(*)}$ that are not
Cabibbo suppressed. Falk and
Petrov \cite{ref9} recently proposed a new method for
measuring $\gamma$ using the partial rates for $CP$-tagged $B_s$
decays. The angle $\gamma$
can also be measured using the $SU(3)$ relations between $B \to \pi K,
\pi\pi $ decay amplitudes \cite{ref10}.
Although this approach is not theoretically
clean in contrast to the $B \to K D $ strategies using pure tree decays
- it is more promising from experimental point of view.
In Ref. \cite{ref8} it is proposed that $\gamma$ can be determined
using the $B \to D^*V~ (V=K^*, \rho)$ modes.

The decays of $B_c$ meson ($\bar cb$ and $\bar b c $ bound states) seem
to be another valuable window for probing the origin of $CP$ violation.
Since large number of $B_c$ meson is expected to be produced at hadronic
colliders like LHC or Tevatron, it is therefore interesting to examine
the features of CP violation in $B_c$ mesons.
In this paper we would like
to discuss about the determination of angle $\gamma$ from the 
pure tree nonleptonic decay modes
$B_c^{\pm} \to D_s^{\pm} \{D^0, \bar D^0, D_{CP}^{\pm}\}$ and the vector
vector modes $B_c^{\pm} \to D_s^{* \pm} D^{*0}(\bar D^{*0}) $.
In Ref. \cite{ref3} the analogous pseudoscalar decay modes
i.e., $B^\pm \to D^0 (\bar D^0) K^\pm $, are
considered for the determination of the angle $\gamma$. However, the
corresponding $B_c$ counterpart has some additional advantages. The isospin
analysis done by Deshpande and Dib \cite{pande01} for the former case
implies that as $D$ and $K$ are isospin 1/2 objects, a $DK $
final state can be either $I=0$ or 1. Since strong interactions conserve
isospin, in general there will be two strong rescattering phases $\delta_0$
and $\delta_1$ , one for each final state of given isospin. However, in case
of $B_c^{\pm} \to D_s^{\pm} D^0 ( \bar D^0)$ modes, the initial
$B_c$ state is an isosinglet and the final states $ D_s^{\pm} D^0 
(\bar D^0)$ are isodoublets. Hence these processes are described by
the effective Hamiltonian of $|\Delta I|=1/2 $. Since isospin is conserved
in strong interactions in general one can expect 
same  strong phases for all the final states.
Hence the uncertainties due
to the presence of the notorious strong phases can be eliminated
without any additional assumptions and $\gamma $ can be 
determined cleanly.
But in actual practice there could be some amount of 
strong phase difference between these  amplitudes  from resonance
effects \cite{falk99}, which we are not taking into
account in this analysis. It has been shown
recently by Fleischer and Wyler \cite{ref11} that 
the $B_c$ counterpart of $B^\pm \to K^\pm D $ triangle approach be well
suited to extract the angle $ \gamma$ as both the amplitudes
$B_c^+ \to D_s^+ D^0$ and $B_c^+ \to D_s^+ \bar D^0$ are of the same order
of magnitude.

The paper is organised as follows. We present the method for
the determination of the angle $\gamma $ from the decay mode $B_c^\pm \to
D_s^\pm D^0 (\bar D^0) $ in section II and
from the vector vector modes $B_c^\pm \to D_s^{* \pm}
D^{*0} (\bar D^{* 0})$ in section III.
Section IV contains our conclusion.

\section{$\gamma$ from $B_c^\pm \to D_s^{\pm}D^0 (\bar{D}^0)$}

The effective Hamiltonians for the decay modes $B_c^- \to D_s^-  D^0$ and
 $B_c^- \to D_s^- \bar D^0$, described by the quark
 level transitions $ b \to c \bar u s$ and $b \to u \bar c s $ respectively
are given as
\begin{eqnarray}
{\cal H}_{eff}(b \to c \bar u s)&=& \frac{G_F}{\sqrt 2} V_{cb} V_{us}^*
\left [ C_1(m_b) ( \bar s u) (\bar c b) + C_2(m_b) (\bar c u)(\bar s b)
\right ] \nonumber\\
{\cal H}_{eff}(b \to u \bar c s)&=& \frac{G_F}{\sqrt 2} V_{ub} V_{cs}^*
\left [ C_1(m_b) ( \bar s c) (\bar u b) + C_2(m_b) (\bar u c)(\bar s b)
\right ]\label{eq:eqn10}
\end{eqnarray}
where $C_1$ and $C_2$ are the Wilson coefficients with values evaluated at
the $b$-quark mass scale as \cite{ref12}
\begin{equation}
C_1(m_b)=1.13,~~~~\mbox{and}~~~~~~C_2(m_b)=-0.29\;,
\end{equation}
$(\bar c b)= \bar c \gamma_\mu(1-\gamma_5)b$ etc. are the usual 
$(V-A)$ color singlet quark currents. 
The hadronic matrix elements of the four quark
current operators i.e., $\langle D_s D|{\cal H}_{eff} |B_c \rangle $ are
very difficult to evaluate from the first principle of QCD. The usual way to
evaluate these matrix elements for nonleptonic $B$-decays is to
assume some approximation. Here we use the factorization approximation
which factorizes each four quark matrix element into a product of
two elements. In the naive factorization hypothesis only the factorizable
contributions are considered. However, the nonfactorizable amplitudes
which cannot be calculated in the naive factorization approach are
important for understanding the data. So the generalized factorization
approach is assumed where these nonfactorizable contributions are
incorporated in a phenomenological way: they are lumped into the
coefficients $a_1 = C_1+C_2/N_c$ and $a_2=C_2+C_1/N_c$, so that now the
effective coefficients $a_1^{eff}$ and $a_2^{eff}$ are treated as
free parameters and their values can be extracted from the experimental
data. From now onwards we shall denote $a_1^{eff}$ and $a_2^{eff}$
by simply $a_1$ and $a_2$. When applying the effective Hamiltonian and
the generalized factorization approximation to $B_c^- \to D_s^- D^0
(\bar D^0)$ decay, one has to take the possible final state interactions
into account. Since here both the final states i.e.,  $ D_s^- D^0$
and  $ D_s^- \bar D^0$ are isospin $1/2$ states, in general
one can expect that both the amplitudes have same
strong FSI phases.
Thus the amplitudes for these decay modes are given as
\begin{eqnarray}
A(B_c^- \to D_s^- \bar D^0) = \frac{G_F}{\sqrt 2} (V_{ub}V_{cs}^*)
e^{i \delta}(a_1 X+ a_2 Y)\nonumber\\
A(B_c^- \to D_s^-  D^0) = \frac{G_F}{\sqrt 2} (V_{cb}V_{us}^*)
e^{i \delta}( a_2 Y)
\end{eqnarray}
where $\delta $ is the strong FSI phase which is same for both the
processes. $X$ and $Y$ are the factorized hadronic matrix elements
\begin{eqnarray}
X&=&\langle D_s^-|(\bar s c) | 0 \rangle \langle \bar D^0 |(\bar u b)|
B_c^- \rangle \nonumber\\
Y&=&\langle \bar D^0|(\bar u c) | 0 \rangle \langle D_s^- |(\bar s b)|
B_c^- \rangle
\end{eqnarray}
Since $V_{ub}=|V_{ub} |e^{-i \gamma}$ the weak phase difference between
the amplitudes $A(B_c^- \to D_s^- \bar D^0)$ and $A(B_c^- \to D_s^- D^0)$
amounts to $-\gamma$ and there is no strong phase difference between them.
As shown in Ref. \cite{ref11}
both these amplitudes are of the same order of magnitude:
\begin{equation}
r = \frac{|A(B_c^- \to D_s^- \bar D^0 )|}{|A (B_c^- \to D_s^- D^0 )| }
={\cal O}(1)\;.
\end{equation}

Now let us write the amplitudes in a more generalized form i.e., in
terms of their magnitudes, strong and weak phases
\begin{eqnarray}
&&A(B_c^- \to D^0 D_s^-) = A_1 e^{i \delta}\nonumber\\
&&A(B_c^- \to \bar D^0 D_s^-) = A_2 e^{-i \gamma} e^{i \delta}
\label{eq:eqn11}
\end{eqnarray}
where $A_1$ and $A_2$ are the magnitudes of the corresponding amplitudes,
$\delta $ is the strong phase and $\gamma $ is the weak phase.
These forms of the amplitudes give $r= A_2/A_1$. The amplitudes
for the corresponding charge conjugate states are given as

\begin{eqnarray}
&&A(B_c^+ \to \bar D^0 D_s^+) = A_1 e^{i \delta}=
A(B_c^- \to D^0 D_s^-)\nonumber\\
&&A(B_c^+ \to  D^0 D_s^+) = A_2 e^{i \gamma} e^{i \delta} =
e^{2 i \gamma} ~A(B_c^- \to \bar D^0 D_s^-)\label{eq:eqn12}
\end{eqnarray}
The decay rates for the flavor specific states of $D$ mesons are given as
(disregarding the phase space factor)
\begin{eqnarray}
&&\Gamma(B_c^- \to D^0 D_s^-) =
\Gamma(B_c^+ \to \bar D^0 D_s^+) = A_1^2\nonumber\\
&&\Gamma(B_c^- \to \bar D^0 D_s^-) =
\Gamma(B_c^+ \to D^0 D_s^+) = A_2^2
\end{eqnarray}

Now from Eqs. (\ref{eq:eqn11}) and (\ref{eq:eqn12}) we can write
the amplitudes for the decay of $B_c^\pm$ into CP eigen state $D_+^0(=
(D^0 +\bar D^0)/\sqrt 2)$ and $ D_s^\pm$ as 

\begin{eqnarray}
\sqrt 2 A (B_c^+ \to D_+^0 D_s^+) = A(B_c^+ \to D^0 D_s^+)+A(B_c^+
\to \bar D^0 D_s^+)\nonumber\\
\sqrt 2 A (B_c^- \to D_+^0 D_s^-) = A(B_c^- \to D^0 D_s^-)+A(B_c^-
\to \bar D^0 D_s^-) \label{eq:eqn13}
\end{eqnarray}

From the two expressions given in Eq. (\ref{eq:eqn13})
one can construct two triangles with the common side
$A(B_c^+ \to \bar D^0 D_s^+)=A(B_c^- \to  D^0 D_s^-)$ and the angle
$(2 \gamma)$ can be determined. This method is recently described
by Fleischer and Wyler \cite{ref11}. The advantage of this method is
that here all sides of the two triangles are of comparable length
giving rise to nonsquashed triangles.

However, here we proceed in a different manner analogous to
Ref. \cite{ref6}. We consider
the decay of $B_c^\pm $ into both CP even $D_+^0((D^0 + \bar D^0)/\sqrt 2)$
and CP odd $D_-^0((D^0 - \bar D^0)/\sqrt 2)$ alongwith the accompanying
$D_s^\pm $ meson i.e.$ B_c^\pm \to D_{+,-}^0 D_s^\pm $, modes.
The amplitudes for $B_c^\pm \to D_-^0 D_s^\pm $ can be written in the
same form as Eq. (\ref{eq:eqn13}) by changing the
sign of second terms. 
Thus neglecting the small $D^0 - \bar D^0$ mixing we obtain the decay
rates into CP eigen states of final $ D$ meson as (the common phase
space factors
are not taken into account)
\begin{eqnarray}
\Gamma(B_c^\pm \to D_+^0 D_s^\pm) = \frac{1}{2}
\left [ A_1^2 +A_2^2+2A_1A_2 \cos \gamma \right ] \nonumber\\
\Gamma(B_c^\pm \to D_-^0 D_s^\pm) = \frac{1}{2}
\left [ A_1^2 +A_2^2-2A_1A_2 \cos \gamma \right ]
\end{eqnarray}
Now we define two charge-averaged ratios for the two CP
eigenstates
\begin{eqnarray}
R_i&=&2\; \frac{\Gamma(B_c^+ \to D_i^0 D_s^+)+\Gamma(B_c^- \to D_i^0 D_s^-)}
{\Gamma(B_c^+ \to \bar D^0 D_s^+)+\Gamma(B_c^- \to D^0 D_s^-)}\nonumber\\
&=&1+ r^2 \pm 2 r \cos \gamma\;,~~~~{\mbox{where}}
~~~~~ i=+,-\label{eq:eqn14}
\end{eqnarray}
This equation can be written in a more generalized form as
\begin{equation}
R_{+,-}=\sin^2 \gamma+(r\pm \cos \gamma)^2
\end{equation}
from which one may get the constraint
\begin{equation}
\sin^2 \gamma \leq R_{+,-}\;.
\end{equation}
The weak phase $\gamma$ can be written in terms of $R_+$ and
$R_-$ from Eq.  (\ref{eq:eqn14}) as
\begin{equation}
\cos \gamma = \frac{1}{4}\frac{R_+ -R_-}{\sqrt{\frac{1}{2}(R_++R_-)-1}}\;.
\end{equation}
Thus the unknown $\gamma$
can be easily determined in terms of $R_+$ and
$R_-$, of course with four
fold quadrant ambiguities. The CP even (odd ) state can be identified
by its CP even (odd) decay products. For instance the states $K_s \pi^0$,
$K_s \rho $, $K_s \omega$, $K_s \phi$ can be used to identify $D_-^0$
while $\pi^+ \pi^-$, $K^+K^-$ represent a $D_+^0$.

The advantage of these decay modes $B_c^{\pm} \to D_s^{\pm}
\{ D^0, \bar{D}^0, D_+^0, D_-^0 \}$ is that there is no FSI strong phase
difference in these decay modes since all these modes have only the
isospin 1/2 final states. So these modes can in principle be considered
as gold plated modes for the extraction of angle $\gamma$. As discussed
by Fleischer and Wyler [20] one expects a huge number of $B_c$ mesons, about
$10^{10}$ untriggered $B_c$'s per year of running and expects
around 20 events per year at LHC for an overall
efficiency of $10\%$. Hence it seems $B_c$ system may well contribute to our
understanding of CP violation.

\section{$\gamma$ from $B_c^\pm \to D_s^{* \pm}D^{*0} (\bar{D}^{*0})$}

The angular distribution for $B_c^- \to D_s^{*-} D^{* 0} (\bar D^{*0})
\to (D_s^- \gamma)(D^0 (\bar D^0) \pi)$ which is same as $B_c^\pm \to
V_1 V_2 \to (l^+ l^-) (P \pi) $ \cite{kramer95} is given
in the
linear polarization basis as \cite{dighe96}
\begin{eqnarray}
&&\frac{1}{\Gamma} \frac{d^3 \Gamma}{d \cos \theta~ d \cos \psi~ d \phi}
= \frac{9}{32 \pi} \biggr[ 2 |A_0|^2 \cos^2 \psi(1-\sin^2 \theta \cos^2
\phi)\nonumber\\
&&+ \sin^2 \psi \left \{ |A_\parallel|^2(1-\sin^2 \theta \sin^2 \phi)
+|A_{\perp}|^2 \sin^2 \theta -\mbox{Im}(A_\parallel^*
A_{\perp}) \sin 2 \theta
\sin \phi \right \} \nonumber\\
&&+ \frac{1}{\sqrt 2} \sin 2 \psi \left \{ {\mbox{Re}} (A_0^* A_{\parallel})
\sin^2 \theta~ \sin 2 \phi + {\mbox{Im}}(A_0^* A_{\perp}) \sin 2 \theta
\cos \phi \right \}\biggr]\label{eq:eqn1}
\end{eqnarray}
where $A_\perp $ is the $P$ wave decay amplitude, $A_0$ and $A_{\parallel}$
are two orthogonal combinations of the $S$ and $D$ wave amplitudes with the
normalization $|A_{\perp}|^2+|A_0|^2 +|A_\parallel|^2=1$.
The angular distribution of $B_c^+ \to D_s^{*+}
D^{*0}(\bar D^{*0}) $ decay is similar to Eq. (\ref{eq:eqn1}) and we shall
use $\bar A $ to indicate the corresponding decay amplitudes. Each amplitude
$A_i$ has both CP conserving FSI phase $\delta_i$ and CP violating weak phase
$\sigma_i $ i.e. $A_i = |A_i| e^{i(\delta_i+\sigma_i)}$ while the
corresponding amplitudes $\bar A_i$ are related to $A_i$ as
\begin{equation}
\bar A_\perp =-|A_\perp| e^{i(\delta_\perp -\sigma_\perp)}\;,~~~~
\bar A_\parallel =|A_\parallel|e^{i(\delta_\parallel -\sigma_\parallel)}\:,
~~~~\mbox{and}~~~\bar A_0=|A_0| e^{i(\delta_0 -\sigma_0)}\;.\label{eq:eqn2}
\end{equation}
The rich kinematics of the vector-vector final state allows one to separate
each of the six combinations of the linear polarization amplitudes 
of Eq. (\ref{eq:eqn1}). The weight factors
for the corresponding amplitudes can be
determined as done in Ref. \cite{ref8} for $B \to V_1 V_2 \to (P_1 \pi)
(P_2 \pi)$ using the Fourier transform in $\phi$ and orthogonality
of Legendre polynomial in $\cos \theta $ and $\cos \psi$. An observable
can be determined from its weight factor $W_i$, given in table-1, using
\begin{equation}
O_i = \frac{32 \pi}{9 }\int d \cos \theta~ d \cos \psi~ d \phi \frac{W_i}
{\Gamma} \frac{d^3 \Gamma}{d \cos \theta~ d \cos \psi~ d \phi} \;.
\end{equation}
It should be noted that in this case the weight factors do not give
identical results under the interchange of $\theta \leftrightarrow
\psi $ as in the case of Ref. \cite{ref8}.
The amplitudes for $B_c^-$ decays for a given polarization state is given
as
\begin{eqnarray}
&&A^\lambda(B_c^- \to D_s^{* -} D^{* 0})=|V_{cb} V_{us}^*|~ a_1^\lambda
~e^{\delta^\lambda}\nonumber\\
&& A^\lambda(B_c^- \to D_s^{* -}\bar{D}^{* 0})=|V_{ub} V_{cs}^*|~
 a_2^\lambda~e^{\delta^\lambda} e^{-i \gamma}
\end{eqnarray}
Since both the final states $D_s^{*-} D^{*0}$ and $D_s^{*-}\bar D^{* 0}$
have isospin 1/2, we have taken same strong FSI phase $\delta^\lambda$
for both the decay modes. The amplitudes for the corresponding charge
conjugate modes can be obtained using Eq. (\ref{eq:eqn2}), as
\begin{eqnarray}
&&A^\lambda(B_c^+ \to D_s^{* +}\bar D^{* 0})=x^\lambda
|V_{cb}^* V_{us}|~ a_1^\lambda~e^{i\delta^\lambda}\nonumber\\
&&A^\lambda(B_c^+ \to D_s^{*+ } D^{* 0})=x^\lambda
|V_{ub}^* V_{cs}|~ a_2^\lambda~e^{i\delta^\lambda} e^{i \gamma}
\end{eqnarray}
where $x^\lambda=-1$ for $\lambda = \perp$ and +1 for
$\lambda=\parallel$ and 0.
We now consider the decay of $D^{*0}/\bar D^{*0} $ into $D^0 \pi^0/
\bar D^0 \pi^0 $ with $D^0/\bar D^0$ meson further decaying to a common
final state $f$. Generally for $B$ decays $f$ is chosen to be a Cabibbo
allowed mode for $\bar D^0$ while it is doubly Cabibbo suppressed
for $D^0$ because in that case the ratio of the amplitudes
$|A(B^- \to \bar D^0 K^-)|/|A(B^- \to D^0 K^-)| \approx {\cal O}(0.1) $.
However, in
$B_c$ decays the ratios of the amplitudes is ${\cal O}(1)$, so here we take
two possible cases : one as done in $B$ case i.e. $D^0 \to f$ is doubly
Cabibbo suppressed while $\bar D^0 \to f $ is Cabibbo allowed and the other
with $f$ as a CP eigen state.

Let us first consider the first case where $f$ can be taken as $K^+ \pi^-$.
Neglecting the negligible mixing effects in the $D^0 - \bar D^0 $ system
the amplitudes for the decays of $B_c^\pm $ to the final state $f$ and its
CP conjugate can be written as

\begin{eqnarray}
A_f^\lambda &=& A^\lambda(B_c^- \to [[f]_D \pi]_{D^*} D_s^{* -} )
\nonumber\\
&=& \sqrt B e^{i \delta^\lambda}\left [|V_{ub} V_{cs}^*|
~a_2^\lambda~e^{-i \gamma} + |V_{cb} V_{us}^*|~R~ a_1^\lambda~
e^{i\Delta}\right ]\nonumber\\
\bar A_{\bar f}^\lambda &=&
A^\lambda(B_c^+ \to [[f]_D \pi]_{D^*} D_s^{* +} )\nonumber\\
&=& x^\lambda \sqrt B  e^{i \delta^\lambda}\left
[|V_{ub}^* V_{cs}|~ a_2^\lambda~e^{i \gamma} +
|V_{cb}^* V_{us}|~R~ a_1^\lambda e^{i\Delta}\right ]\nonumber\\
A_{\bar f}^\lambda &=&A^\lambda(B_c^- \to [[\bar f]_D
\pi]_{D^*} D_s^{* -} )\nonumber\\
&=&\sqrt B e^{i \delta^\lambda} \left [|V_{ub} V_{cs}^*|
~R~  a_2^\lambda~
e^{-i \gamma}e^{i \Delta} + |V_{cb} V_{us}^*|~ a_1^\lambda~
\right ]\nonumber\\
\bar A_f^\lambda &=&
A^\lambda(B_c^+ \to [[f]_D \pi]_{D^*} D_s^{*+} )\nonumber\\
&=&x^\lambda \sqrt B e^{i \delta^\lambda}
\left [|V_{ub}^* V_{cs}|~R~ a_2^\lambda~
e^{i \gamma} e^{i \Delta} + |V_{cb}^* V_{us}|~ a_1^\lambda~
\right ]\label{eq:eqn01}
\end{eqnarray}
where $[X]_M$ indicates that the state $X$ is constructed to have
invariant mass of $M$; $B=Br(\bar D^0 \to f)$, $R^2 =Br(\bar D^0 \to f)/
Br(D^0 \to f)$ and $\Delta $ is the strong phase difference between
$\bar D^0 \to f$ and $\bar D^0 \to \bar f$.

Thus the measurement of the angular distribution for each of the four
modes provides us with a total of twentyfour observables, six for each mode.
These observables can be extracted experimentally using the weight
functions. There are only thirteen unknowns in Eq.(\ref{eq:eqn01}) :
$R$, $\Delta$,
$\gamma$, $|V_{ub}|$ and three variables for each $a_1^\lambda$,
$a_2^\lambda$ and $\delta^\lambda$. Thus $\gamma$ in principle can be 
easily determined from these observables.

Now let us consider $f$ to be a CP eigenstate i.e. $f= K^+ K^-$
or $ \pi^+ \pi^-$ with CP eigenvalue +1. In this case the number of
unknowns in Eq. (\ref{eq:eqn01})
is further reduced because there is no relative
strong phase difference between $\bar D^0 \to f$ and $\bar D^0 \to
\bar f$
as $f$ is a CP eigen state. So there will no longer be the strong
phase difference factor $(e^{i\Delta} )$ in the expressions for the
amplitudes (\ref{eq:eqn01}).
Furthermore, since $f$ is chosen to be a CP eigenstate
$R$ is also no longer  an unknown. It can be related to $Br (\bar D^0 \to
f)$ through the experimentally determined CP rate asymmetries $a_{CP}(f)$.
Defining $a_{CP}(f)$ as
\begin{equation}
a_{CP}(f)= \frac{Br(\bar D^0 \to f)-Br(D^0 \to f)}
{Br(\bar D^0 \to f)+Br(D^0 \to f)}
\end{equation}
which gives
\begin{equation}
R=\frac{1-a_{CP}(f)}{1+a_{CP}(f)}
\end{equation}
Since $a_{CP}$ is very small i.e., $ a_{CP}(K^+K^-)=0.026 \pm  0.03$ 
and $ a_{CP}(\pi^+\pi^-)=-0.05 \pm  0.08$ \cite{pdg00}, $R$ can be taken as
approximately 1.

Thus we get rid of two more unknowns $R$ and $\Delta$ if we consider
the common final state $f$ to be a CP eigenstate. In this case the total
number of unknowns are eleven which would possibly
be overdetermined with the
twentyfour observables.

Now let us consider whether the decay modes $B_c^{\pm} \to D_s^{* \pm}
D^{ *0}(\bar D^{* 0})$ can be used to observe CP violation or not.
It is well known that to observe direct CP violation one would
require two interferring amplitudes with different strong and weak
phases. Thus the usual signature of CP violation i.e. the partial
rate asymmetry
\begin{equation}
|A_f^\lambda|^2 - |\bar A_{\bar f}^\lambda|^2 =4 |V_{ub}^* V_{cs}V_{cb}
V_{us}^*|~ R~ B~ a_1^\lambda~ a_2^\lambda~ \sin \Delta \sin \gamma\;,
\end{equation}
can not be observed in case $f$ is chosen 
to be a CP eigen state.
It is therefore interesting to see if there are other observables in the
angular distribution which can provide useful information about CP
violation even if partial rate asymmetries are zero. It is clear that
the coefficients $\hat\alpha = -{\rm{Im}}(A_{\parallel}^* A_\perp)$,
$\hat\gamma = {\rm{Im}}(A_{0}^* A_\perp)$, the fourth and last terms in
the expression for
angular distribution (\ref{eq:eqn1}), and similarly $\bar{\hat\alpha} $
and $\bar{\hat\gamma} $ for $B_c^+$ decay,
contain information about CP violation,
even for cases where the two amplitudes have no relative strong phase
difference. From Eq. (\ref{eq:eqn01})
we can obtain the following quantities
which can measure CP violation as
\begin{eqnarray}
&&{\rm{Im}}\{(A^\perp A^{\rho *})_f +(\bar A^\perp \bar A^{\rho *})_
{\bar f}\}\nonumber\\
 &&=2 |V_{ub}^* V_{cs}V_{cb}V_{us}^*| R B \sin \gamma [a_1^\perp a_2^\rho
 \cos(\delta^\perp - \delta^\rho+\Delta)
 -a_2^\perp a_1^\rho \cos(\delta^\perp - \delta^\rho-\Delta)]\nonumber\\
 &&{\rm{Im}}\{(A^\perp A^{\rho *})_{\bar f}+(\bar A^\perp
 \bar A^{\rho *})_{f}\}\nonumber\\
& &=2 |V_{ub}^* V_{cs}V_{cb}V_{us}^*| R B \sin \gamma [a_1^\perp a_2^\rho
 \cos(\delta^\perp - \delta^\rho-\Delta)
 -a_2^\perp a_1^\rho \cos(\delta^\perp - \delta^\rho+
 \Delta)]\nonumber\\
&&  {\rm{Im}}\{(A^\perp A^{\rho *})_f +(\bar A^\perp \bar A^{\rho *})_{\bar f}
 +(A^\perp A^{\rho *})_{\bar f}+
 (\bar A^\perp \bar A^{\rho *})_{f}  \}\nonumber\\
& &=4 |V_{ub}^* V_{cs}V_{cb}V_{us}^*| R B \sin \gamma
  \cos(\Delta) \cos(\delta^\perp - \delta^\rho)
 [a_1^\perp a_2^\rho
 -a_2^\perp a_1^\rho ]\;,
 \end{eqnarray}
 where $\rho= \parallel$ or 0. When $\rho=\parallel$ we observe CP
 violation in $\hat\alpha$ parameter and $\rho=0$ corresponds to
 $\hat\gamma$
 asymmetries. These CP violating observables do not require FSI
 phase differences and are especially sensitive to CP violating weak phases.
 Thus unlike $B_c^\pm \to D_s^\pm D^0 (\bar D^0)$ decays where direct CP
 violation can not be observed, whereas in this case
 one can observe the signatures of CP violation. However, the
 weak phase $\gamma$ can be cleanly
 determined in both the cases.

Now let us make a crude estimate of the number of events required to
measure $\gamma $ by this method, assuming that about $10^{10}$
untriggered $B_c$'s will be available at LHC per year of running. The
observed number of events for each mode is \cite{ref7}
\begin{equation}
N_{obs}=N_{0} \times Br \times f \times \epsilon\;,
\end{equation}
where $N_{0}$, $Br$, $f$ and $\epsilon$ are the total
number of $B_c$ events, branching ratio, observation  fraction
and detector efficiency, respectively.
The particle $\bar D^0 $ is seen in its flavor tagging $K^+ \pi^-$
and $K^+ \pi^- \pi^+ \pi^- $ modes. Ref. \cite{ref4} gives the list of
visible fraction ($f$) and the detector efficiencies $(\epsilon )$ for
the various final state particles. 
The decay width for the color
suppressed mode $B_c^+ \to D_s^{*+} D^{*0} $ is estimated in Ref.
\cite{chang94} with value
$\Gamma(B_c^+ \to D_s^{*+} D^{*0} )= a_2^2~0.564 \times 10^{-13}$ MeV.
Using $a_2=0.23$, the branching ratio is found to be
\begin{equation}
Br(B_c^+ \to D_s^{*+} D^{*0} )= 2 \times 10^{-6}\;.
\end{equation}
For the decay mode  $B_c^+ \to D_s^{*+} \bar D^{*0}$ we expect a branching
ratio at $10^{-5}$ level.

If we assume the visible fraction ($f$) of the final $D$ meson
to be $11.5\%$ and al overall efficiency of $10 \%$ \cite{ref4}
we get approximately
230 reconstructed events. This crude estimate indicates that the $B_c$
system may well contribute to our understanding of CP violation. 

\section{Conclusion}
In this paper we have discussed the determination of the angle $\gamma$
from the pure tree nonleptonic $B_c$ decay modes $B_c^\pm \to D_s^\pm D^0
(\bar D^0)$ and $B_c^\pm \to D_s^{* \pm} D^{* 0}
(\bar D^{* 0})$. For the former case we have followed the method of Gronau
\cite{ref6}.
However, the decay modes  $B_c^\pm \to D_s^\pm D^0 (\bar D^0)$
are particularly interesting due to the following reasons.
First they are free from the presence of the relative strong phases
between $B_c^+ \to D_s^+ D^0 $ and
$B_c^+ \to D_s^+ \bar D^0 $ as they are both isospin 1/2 states. Secondly
the two interferring amplitudes are of the same order of magnitude i.e.,
the ratio of their amplitudes $r = |A(B_c^- \to D_s^- \bar D^0 )|/
|A(B_c^- \to D_s^- D^0)| \approx {\cal O}(1)$ whereas for the corresponding
analog system $ B^\pm \to D K^\pm $, $r$ is ${\cal O}(0.1)$. 

For the vector-vector final states we have followed the method of
Ref \cite{ref8}. Because of no relative strong phase 
between the two interferring
amplitudes the number of unknowns are much less than the number of
observables. Since the magnitudes of the amplitudes are of the same order
here we have considered two possibilities for the common final
state $f$ ($ D^{*0}/ \bar D^{* 0} \to D^0 \pi^0/ \bar D^0 \pi^0 \to
f \pi^0/ f \pi^0 $).
({\it i}) $f$ is a Cabibbo allowed mode for $\bar D^0 $ and hence
doubly Cabibbo suppressed for $D^0$. ({\it ii}) $f$ is a $CP$
eigen state. Due to the rich kinematics of vector-vector final
states in this case one can observe alternative signature of $CP$
violation even when there is zero partial rate asymmetry.

To summarize,  it is possible to determine the
weak phase $\gamma$ from the measurement of the nonleptonic decay
modes $B_c^\pm \to D_s^\pm D^0(\bar D^0)$ and the corresponding vector
vector modes $B_c^\pm \to D_s^{* \pm} D^{*0}(\bar D^{*0})$ cleanly
without any hadronic uncertainties (since they
are free from the presence of
relative strong phase difference and  penguin pollutions).
Further, in the vector-vector
modes the determination is even cleaner as the number of
observables is much more than the number of unknowns.
Hence, these decay modes can in principle be considered as gold plated modes
for the determination of the angle $\gamma$.

\section{Acknowdgements}
AKG would like to thank Council of Scientific and Industrial Research,
Government of India, for financial support.

\begin{table}
\caption{ The weight factors corresponding to the observables
in the angular distribution for $B_c \to D_s^* D^{*0} $ decay modes.
}
\vspace {0.2 true in}
\begin{tabular}{|c|c|}
\hline
& \\
\multicolumn{1}{|c|}{Observable}&
\multicolumn{1}{|c|}{ Weight Factor} \\
& \\
\hline
& \\ 
$|A_0|^2 $ & $\frac{9}{64 \pi} \left (5 \cos^2\psi -1\right )$\\ 
& \\
$|A_\parallel|^2$& $\frac{9}{64 \pi} \left (8 \cos^2\phi -
5 \sin^2 \theta\right )$\\
 &  \\
$|A_\perp|^2 $& $\frac{9}{32 \pi} \left (2-5 \cos^2\theta \right )$\\ 
&\\ 
$\mbox{Im}(A_\parallel^* A_\perp ) $
&$ \frac{15}{\pi^2} \cos^2 \psi \cos \theta \sin
\phi $\\
& \\
$\mbox{Re}(A_0^*A_\parallel)$
& $\frac{3}{\pi^2}\sqrt{2} \cos \psi \sin
2\phi $\\
&\\
$\mbox{Re}(A_0^* A_\perp ) $
& $\frac{16}{\pi^3} \sqrt{2}\cos \psi \cos \theta \cos
\phi $\\
\hline
\end{tabular}
\end{table}

\end{document}